\documentclass[12pt]{article}
\usepackage{longtable,supertabular}
\usepackage{amsmath}
\usepackage{amssymb}
\usepackage{latexsym}
\usepackage{cite,a4wide,color}
\usepackage{multirow}
\usepackage{ntheorem}
\usepackage[square,numbers]{natbib}

\theoremstyle{remark}

\newcommand{\bC}{{\mathbf{{C}}}}
\newcommand{\bF}{{\mathbf{{F}}}}

\newcommand{\bc}{{\mathbf{c}}}

\newcommand{\bv}{{\mathbf{v}}}

\newcommand{\Plotkin}{{\mathrm{Plotkin}}}

\title{\bf  Quasi-Perfect and Distance-Optimal Sum-Rank Codes}
\author{Hao Chen \thanks{Hao Chen is with the College of Information Science and Technology/Cyber Security, Jinan University, Guangzhou, Guangdong Province, 510632, China (e-mail: haochen@jnu.edu.cn).
		The research of Hao Chen was supported by NSFC Grant 62032009.
}}

\begin{document}
	
	\maketitle
	\begin{abstract}
		Constructions of quasi-perfect and distance-optimal codes are interesting and challenging problems. In this paper, we give the following three results.\\

1) If $\lambda|q^{sm}-1$ and $\lambda <\sqrt{\frac{(q^s-1)}{2(q-1)^2(1+\epsilon)}}$, an infinite family of distance-optimal  $q$-ary sum-rank codes with the block length $t=\frac{q^{sm}-1}{\lambda}$, the matrix size $s \times s$, the cardinality $q^{s^2t-s(2m+3)}$ and the minimum sum-rank distance four is constructed.\\

2) Block length $q^4-1$ and the matrix size $2 \times 2$ distance-optimal sum-rank codes with the minimum sum-rank distance four and the Singleton defect four are constructed. These sum-rank codes are close to the sphere packing bound , the Singleton-like bound and have much larger block length $q^4-1>>q-1$.\\

3) For given positive integer $m$ satisfying $2\leq m$, an infinite family of quasi-perfect sum-rank codes with the matrix size $2 \times m$, and the minimum sum-rank distance three is also constructed. Families of quasi-perfect binary sum-rank codes with the matrix size $2 \times 2$ and the minimum sum-rank distance four are also presented. These sum-rank codes are distance-optimal automatically.\\

Almost MSRD $q$-ary codes with the block lengths up to $q^2$ are presented. We also show that new distance-optimal binary sum-rank codes can be obtained from the Plotkin sums.\\

 Distance-optimal sum-rank code. Singleton defect. Almost MSRD code. Plotkin sum of sum-rank codes.

	\end{abstract}
	
	\newpage

\section{Introduction}

\subsection{Preliminaries}

Sum-rank codes have found wide applications in multishot network coding, see \cite{MK19,NPS,NU}, space-time coding, see \cite{Lu,SK}, and coding for distributed storage,  see \cite{MK,MP1}. There have attracted many attentions in recent years, see e.g.\cite{BGR,MP1,MP191,MK19,MP21,Neri,NSZ21} and references therein. Fundamental properties and some bounds on sizes of sum-rank codes were given in the paper \cite{BGR}. For a nice survey of sum-rank codes and their applications, we refer to \cite{MPK22}.\\

Let ${\bf F}_q^{(u,v)}$ be the set of all $ u \times v$ matrices. Let $u_i \leq v_i$ be $2n$ positive integers satisfying $v_1 \geq v_2 \cdots \geq v_n$. Set $N=u_1+\cdots+u_n$.  Let $${\bf F}_q^{(u_1, v_1), \ldots,(u_n, v_n)}={\bf F}_q^{u_1 \times v_1} \bigoplus \cdots \bigoplus {\bf F}_q^{u_n \times v_n}$$ be the set of all ${\bf x}=({\bf x}_1,\ldots,{\bf x}_t)$, ${\bf x}_i \in {\bf F}_q^{u_i \times v_i}$, $i=1,\ldots,n$. This is a linear space over ${\bf F}_q$ of the dimension $\Sigma_{i=1}^n u_iv_i$. We call $u_1 \times v_1, \ldots, u_n \times v_n$,  matrix sizes of sum-rank codes. Set $wt_{sr}({\bf x}_1, \ldots, {\bf x}_n)=rank({\bf x}_1)+\cdots+rank({\bf x}_n)$ and $$d_{sr}({\bf x},{\bf y})=wt_{sr}({\bf x}-{\bf y}),$$ for ${\bf x}, {\bf y} \in {\bf F}_q^{(u_1,v_1), \ldots,(u_n,v_n)}$. This is indeed a metric on ${\bf F}_q^{(u_1,v_1), \ldots,(u_n,v_n)}$.  This sum-rank metric satisfies $$d_{sr}({\bf x}+{\bf y}, {\bf z}+{\bf y})=d_{sr}({\bf x}, {\bf y}). $$ Then numbers of elements in balls with a given radius in the sum-rank metric space is independent of their centers. Let $V_{sr}(q, r)$ be the number of elements in a ball with the radius $r$, see \cite{BGR,MPK22}.\\

A $q$-ary sum-rank code ${\bf C}$ of block length $n$ and matrix sizes $u_1 \times v_1, \ldots, u_n\times v_n$ is a subset of  the finite metric space ${\bf C} \subset {\bf F}_q^{(u_1,v_1), \ldots,(u_n, v_n)}$. Its minimum sum-rank distance is defined by $$d_{sr}({\bf C})=\min_{{\bf x} \neq {\bf y}, {\bf x}, {\bf y} \in {\bf C}} d_{sr}({\bf x}, {\bf y}).$$  When ${\bf C} \subset {\bf F}_q^{(u_1,v_1), \ldots,(u_n,v_n)}$ is a linear subspace over ${\bf F}_q$, ${\bf C}$ is called a linear sum-rank code. When $q=2$, this code is called a binary sum-rank code. When $u_i=v_i=1$, $i=1, \ldots, n$, error-correcting codes in the Hamming metric are just sum-rank codes with the matrix size $1 \times 1$.\\

The Hamming weight $wt({\bf a})$ of a vector ${\bf a}=(a_0, \ldots, a_{n-1}) \in {\bf F}_q^n$ is the cardinality of its support, $$supp({\bf a})=\{i: a_i \neq 0\}.$$ The Hamming distance $d({\bf a}, {\bf b})$ between two vectors ${\bf a}$ and ${\bf b}$ is $d({\bf a}, {\bf b})=wt({\bf a}-{\bf b})$. Then ${\bf F}_q^n$ is a finite Hamming metric space. The minimum (Hamming) distance of a code ${\bf C} \subset {\bf F}_q^n$ is, $$d=\min_{{\bf a} \neq {\bf b}} \{d({\bf a}, {\bf b}),  {\bf a} \in {\bf C}, {\bf b} \in {\bf C} \}.$$  The Hamming metric is just the sum-rank metric on the space $\bF_q^{(1,1)} \oplus \cdots \oplus \bF_q^{(1,1)}=\bF_q^n$. One of the main goal of theory of error-correcting codes is to construct codes ${\bf C} \subset {\bf F}_q^t$ with large cardinalities and minimum distances. In general, there are some upper bounds on cardinalities or minimum distances of codes. Optimal codes attaining these bounds are particularly interesting, see \cite{MScode}.\\

A linear $q$-ary code with the length $n$, the dimension $k$ (and the minimum distance $d$), is denoted by a linear $[t,k]_q$ (or $[n,k,d]_q$) code. A $q$-ary code with the length $t$, the cardinality $M$, and the minimum distance $d$, is denoted by an $(n, M, d)_q$ code. The Singleton bound for general codes asserts $M \leq q^{n-d+1}$ and codes attaining this bound is called (nonlinear) maximal distance separable (MDS) codes. Reed-Solomon codes are well-known linear MDS codes, see \cite{MScode}. Construction of non-Reed-Solomon type linear MDS codes have attracted many attentions, see, e.g.,  \cite{Chen} and references therein. Cyclic and Bose-Chaudhuri-Hocquenghem (BCH) codes are one of the most important class of codes in theory and practice, see \cite{MScode,DingLi}.\\

For a code ${\bf C}$ in the Hamming metric space ${\bf F}_q^n$, we define
its covering radius  by $$R_{covering}({\bf C})=\max_{{\bf x} \in {\bf F}_q^n} \min_{{\bf c}
\in {\bf C}} \{wt({\bf x}-{\bf c})\}.$$ Hence the Hamming balls $B(x, R_{covering}({\bf C}))$
centered at all codewords $x \in {\bf C}$, with the radius $R_{covering}({\bf  C})$ cover the
whole space ${\bf F}_q^n$. Then ${\bf C}$ is called a covering code with the radius $R_{covering}({\bf C})$.
We refer to the excellent book \cite{CHLL} on this classical topic in coding theory. \\

A code satisfying $$R_{covering}({\bf C})=\lfloor \frac{d({\bf C})-1}{2} \rfloor$$ is perfect, see \cite{Lint,MScode,CHLL}.  It was proved that perfect codes have the same parameters as Hamming codes and Golay codes, see \cite{Lint,CHLL}. We refer to \cite{Etzion1} for nonlinear perfect codes.  The next best possibilities are quasi-perfect codes. A code ${\bf C}$ satisfying $$R_{covering}({\bf C})=\lfloor \frac{d({\bf C})-1}{2}\rfloor+1$$ is called quasi-perfect, see \cite[Section 5, Chapter 1]{MScode} and \cite{Wagner,Etzion,Etzion1}. Quasi-perfect codes are obviously ideal candidates for these parameters with which there is no perfect code.\\

The sphere packing bound for $(n, M, d)_q$ codes asserts that $$M \cdot V_H(q, \lfloor \frac{d-1}{2}\rfloor) \leq q^n,$$ where $V_H(q, r)=1+n(q-1)+\displaystyle{n \choose 2}(q-1)^2+\cdots+\displaystyle{n \choose r}(q-1)^r$ is the volume of the ball with the radius $r$ in the Hamming metric space ${\bf F}_q^n$, since balls centered at codewords with the radius $\lfloor \frac{d-1}{2} \rfloor$ are disjoint, see \cite{MScode}. If there is an $(n, M, d)_q$ code, and there is no $(n, M, d+1)_q$ code, this $(n, M, d)_q$ code is called distance-optimal. Some distance-optimal codes are with respect to the sphere packing bound. That means, there is an $(n, M, d)_q$ code and $M \cdot V_H(q, \lfloor \frac{d}{2}\rfloor) >q^n$. For a linear $[n, k, d]_q$ code with the codimension $r=n-k$ satisfying $$V_H(q, \frac{d}{2}) >q^r,$$ this linear $[n, k, d]_q$ code is distance-optimal with respect to the sphere packing bound.\\

The Singleton-like bound for sum-rank codes with the matrix size $u \times v$, $u \leq v$ is $$|{\bf C}| \leq q^{v(ut-d_{sr}+1)}.$$ Similar to the case in Hamming metric, the difference $v(ut-d_{sr}({\bf C})+1)-\log_q |{\bf C}|$ is called the Singleton defect of the sum-rank code ${\bf C}$. A  sum-rank code attaining this bound (Singleton defect zero) is called a maximum sum-rank-metric distance (MSRD) code. When $n \leq q-1$ and $N \leq (q-1)v$, MSRD codes attaining the above Singleton-like bound were constructed in \cite{MP1}. They are called linearized Reed-Solomon codes, we also refer to \cite{Neri} for the further results. We also refer to \cite[Section VII]{BGR} and \cite{AKR} for other results on MSRD codes.  Notice that Singleton defects are $0, v, 2v, 3v, \ldots$. Singleton defect $v$ sum-rank codes with the matrix size $u \times v$ can be called almost MSRD codes. When $u=v=1$, these almost MSRD codes are just almost MDS codes in the Hamming metric. MSRD codes over finite chain rings were constructed in a recent paper \cite{MPP}. Several new constructions of MSRD codes were given in \cite{MP24}.\\

The sphere packing bound for a sum-rank code ${\bf C}$ with $M$ codewords and the minimum sum-rank distance $d_{sr}$ asserts $$M \cdot V_{sr}(q, \lfloor \frac{d_{sr}-1}{2} \rfloor) \leq q^{u_1v_1+\cdots+u_nv_n},$$ see \cite{BGR}. A sum-rank code in ${\bf F}_q^{(u_1, v_1)} \bigoplus \cdots \bigoplus {\bf F}_q^{(u_n,v_n)}$, with the minimum sum-rank distance $d_{sr}$ and $M$ codewords is called distance-optimal, if there is no sum-rank code with the minimum sum-rank distance $d_{sr}+1$ and $M$ codewords. It is obvious if there is a sum-rank code ${\bf C}$ with the minimum sum-rank distance $d_{sr}$ and $M$ codewords, and $$V_{sr}(q, \lfloor \frac{d_{sr}}{2} \rfloor) \cdot M>q^{u_1v_1+\cdots+u_nv_n},$$ this code ${\bf C}$ is distance-optimal with respect to the sphere packing bound. Therefore a linear (over ${\bf F}_q$) sum-rank code with the codimension $r$ is distance-optimal with respect to the sphere packing bound, if $$V_{sr}(q, \lfloor \frac{d_{sr}}{2} \rfloor)>q^r,$$ is satisfied. A sum-rank code ${\bf C} \subset {\bf F}_q^{(u_1,v_1)} \oplus \cdots \oplus {\bf F}_q^{(u_n,m_n)}$  with the minimum sum-rank distance $d_{sr}$ satisfying $$V_{sr}(q,\lfloor \frac{d_{sr}-1}{2} \rfloor) \cdot |{\bf C}|=q^{u_1v_1+\cdots+u_nv_n}$$ is a perfect sum-rank code. The whole space is the disjoint union of balls with the radius $\lfloor \frac{d_{sr}-1}{2} \rfloor$ centered at all codewords of a perfect code ${\bf C}$.\\

Let ${\bf C} \subset {\bf F}_q^{(u_1,v_1), \ldots, (u_n, v_n)}$ be a sum rank code, its covering radius $R_{sr}({\bf C})$ is the smallest radius $R_{sr}$ such that the balls $$B_{sr}({\bf x}, R_{sr})=\{{\bf y} \in {\bf F}_q^{(u_1,v_1), \ldots, (u_n, v_n)}: wt_{sr}({\bf y}-{\bf x}) \leq R_{sr}\},$$ centered at all codewords ${\bf x} \in {\bf C}$,  cover the whole space ${\bf F}_q^{(u_1,v_1), \ldots, (u_n,v_n)}$. Perfect sum-rank codes are just these codes satisfying $$R_{sr}({\bf C})=\lfloor \frac{d_{sr}(\bC)-1}{2} \rfloor.$$ As in the case of the sum-rank metric with the matrix size $1 \times 1$ (the Hamming metric), a sum-rank code satisfying $$R_{sr}({\bf C})=\lfloor \frac{d_{sr}(\bC)-1}{2} \rfloor+1$$ is called quasi-perfect. Quasi-perfect sum-rank codes are naturally extensions of quasi-perfect codes of the matrix size $1 \times 1$ in the Hamming metric. Then how to construct quasi-perfect sum-rank codes with the matrix size $u \times v$, where both $u$ and $v$ are larger than $1$, is an interesting and challenging problem.\\

\subsection{Related works and our contributions}

Some distance-optimal ternary cyclic codes with the length $3^m-1$ and minimum distance four were constructed in \cite{YCD,Ding}. Distance-optimal codes with minimum distance four and six were constructed in \cite{YCD,Ding,Heng,Heng1,Xiong}. There are only few distance-optimal codes with minimum distance six reported in the literature, see \cite{Heng} and \cite[Theorem 9]{Wang}. Several infinite families of optimal codes with respect to the Griesmer bound were constructed in \cite{Hu}. For distance-optimal sum-rank codes, we refer to \cite{MP191,Cheng}. Distance-optimal (actually perfect) sum-rank codes with minimum sum-rank distance three were constructed in \cite{MP191}. Perfect sum-rank codes constructed by U. Mart\'{\i}nez-Pe\~{n}as, in \cite{MP191} have the matrix size $1 \times n$. He asked in \cite[Remark 5]{MP191} if there is perfect sum-rank codes for the matrix size $m \times n$, $m>1, n>1$.  On the other hand, the first infinite family of binary cyclic sum-rank codes with minimum sum-rank distance four was constructed in \cite[Section 7]{Cheng}.\\

The constructions of quasi-perfect codes for various parameters has received many attentions. Many binary, ternary linear or nonlinear quasi-perfect codes and quasi-perfect linear codes over large fields were constructed and classified, see \cite{Dodun,Etzion,Etzion1,LiHell,BBDF,Wagner,Giulietti,LiHell}. In particular, short quasi-perfect codes over ${\bf F}_q$ with the minimum distance four were constructed in \cite{Giulietti}. It is direct to verify that quasi-perfect codes with even minimum distances are distance-optimal. Therefore quasi-perfect codes in \cite{Giulietti} are distance-optimal.\\

The existence and constructions of almost MDS codes and near MDS codes in the Hamming metric have been studied for many years, see e.g., \cite{Boer,HF} and references therein. No almost MSRD codes with large block lengths is reported in the literature.\\

In this paper, we construct infinitely many new families of distance-optimal $q$-ary cyclic codes with the minimum distance four. These families of distance-optimal sum-rank codes have new parameters, compared with previous distance-optimal codes constructions. Infinitely many families of distance-optimal $q$-ary cyclic sum-rank codes with minimum distance four are presented. MSRD codes are sum-rank codes with the zero Singleton defect, these MSRD codes constructed in \cite{MP1,Neri} have the  block lengths at most $q-1$. It is interesting to ask if longer block lengths are required, how about their Singleton defects. Singleton defect two almost MSRD $q$-ary codes with the block length up to $q^2$ and the minimum sum-rank distance four are given. An infinite family of distance-optimal $q$-ary sum-rank codes with the block length $q^4-1$, the matrix size $2 \times 2$ and the minimum sum-rank distance four is constructed.The Singleton defect of these distance-optimal sum-rank codes is $4$.  It is showed that Plotkin sums of sum-rank codes lead to more distance-optimal sum-rank codes. Quasi-perfect binary sum-rank codes of the matrix size $2 \times 2$ and the minimum sum-rank distance four are constructed. Notice that these quasi-perfect sum-rank codes are distance-optimal automatically.\\

More importantly, an infinite family of quasi-perfect $q$-ary sum-rank codes with the matrix size $2 \times m$, where $m$ is a positive integer satisfying $2 \leq m$, and the minimum sum-rank distance three, is constructed. We also construct the matrix size $2 \times 2$ quasi-perfect binary sum-rank codes with the minimum sum-rank distance four from quasi-perfect codes over $\bF_4$ given in \cite{Giulietti}. These are the first several families of quasi-perfect sum-rank codes with the matrix size $n \times m$, where both $m$ and $n$ are larger than $1$.\\

The paper is organized as follows. In Section 2, we recall basic facts about cyclic codes, which are used for constructions in this paper. A basic construction proposed in our previous paper \cite{Chen1} was also included. In Section 3, distance-optimal $q$-ary cyclic codes with the minimum distance four are constructed. In Section 4 and 6, distance-optimal cyclic sum-rank codes with the minimum sum-rank  distance four are constructed. In Section 6, quasi-perfect sum-rank codes with the matrix size $2 \times m$ are constructed. In Section 5, almost MSRD codes with the minimum sum-rank distance four are presented. In Section 7 and 8, quasi-perfect binary sum-rank codes with the minimum sum-rank distance three and four are constructed. In Section 9, Plotkin sums of sum-rank codes are introduced and new distance-optimal binary sum-rank codes are constructed. Section 10 concludes the paper.\\

\section{Cyclic codes and the construction of sum-rank codes from codes in the Hamming metric}

Cyclic codes was introduced by E. Prange in 1957, see \cite{Prange}. They are one of the most important class of linear codes in coding theory and practice. If $(c_0, c_1, \ldots, c_{n-1}) \in {\bf C}$, then $(c_{n-1}, c_0, \ldots, c_{n-2}) \in {\bf C}$, this code ${\bf C} \subset {\bf F}_q^n$ is called cyclic. The dual code of a cyclic code is a cyclic code. A codeword ${\bf c}$ in a cyclic code is identified with a polynomial ${\bf c}(x)=c_0+c_1x+\cdots+c_{n-1}x^{n-1}\in {\bf F}_q[x]/(x^n-1)$. Every cyclic code is a principal ideal in the ring ${\bf F}_q[x]/(x^n-1)$ and then generated by a factor $g$ of $x^n-1$. BCH codes were introduced in 1959-1960, see \cite{BC1,BC2,Hoc}.  Let $n$ be a positive integer satisfying $\gcd(n,q)=1$, and ${\bf Z}_n={\bf Z}/n{\bf Z}=\{0, 1, \ldots, n-1\}$ be the residue classes module $n$. A subset $C_i$ of ${\bf Z}_n$ is called a $q$-cyclotomic coset if $$C_i=\{i, iq, \ldots, iq^{l-1}\},$$ where $i \in {\bf Z}_n$ is fixed and $l$ is the smallest positive integer such that $iq^l \equiv i$ $mod$ $n$. It is clear that $q$-cyclotomic cosets correspond to irreducible factors of $x^n-1$ in ${\bf F}_q[x]$. A generator polynomial of a cyclic code is the product of several irreducible factors of $x^n-1$. The defining set of a cyclic code generated by $g$  is the the following set $${\bf T}_{{\bf g}}=\{i: g(\beta^i)=0\}.$$ Hence the defining set of a cyclic code is the disjoint union of several $q$-cyclotomic cosets. If there are $\delta-1$ consecutive elements in the defining set of a cyclic code, the minimum distance of this cyclic code is at least $\delta$. This is the BCH bound and $\delta$ is the designed distance, see \cite{BC1,BC2,Hoc,MScode}. There have been numerous papers on BCH codes, we refer to the recent survey \cite{DingLi}. Some further bounds on minimum distances of cyclic codes such as the Hartmann-Tzeng bound and the Roos bound were proposed and used to construct good cyclic codes see \cite[Chapter 6]{Lint} and \cite[Chapter 4]{HP}.\\

Cyclic and negacyclic sum-rank codes in ${\bf F}_q^{s \times s} \bigoplus \cdots \bigoplus {\bf F}_q^{s \times s}$ ($n$ copies) were introduced in \cite{Cheng}. A sum-rank code ${\bf C} \subset {\bf F}_q^{s \times s} \bigoplus \cdots \bigoplus {\bf F}_q^{s \times s}$ is called cyclic, if $({\bf x}_0, \ldots, {\bf x}_{n-1}) \in {\bf C}$, then $({\bf x}_{n-1}, {\bf x}_0, \ldots, {\bf x}_{n-2}) \in {\bf C}$. A direct construction of cyclic and negacyclic sum-rank codes from cyclic and negacyclic codes in the Hamming metric was given in \cite{Cheng}. Cyclic skew cyclic sum-rank codes introduced in \cite{MP21} are special cases of cyclic sum-rank codes.\\

We identify the linear space ${\bf F}_q^{(u,u)}$ as the linear space of $q$-polynomials, $${\bf F}_q^{(u,u)}=\{a_0x+a_1x^q+\cdots+a_{u-1}x^{q^{u-1}}:a_i \in {\bf F}_{q^u}, i=0,1,\ldots,u-1\}.$$  Let ${\bf C}_0, \ldots, {\bf C}_{u-1}$ be linear $[n, k_i, d_i]_{q^u}$ codes over ${\bf F}_{q^u}$. The sum-rank codes constructed in \cite{Chen1} is as follows, $$SR({\bf C}_0, \ldots, {\bf C}_{u-1})=\{{\bf a}_0x+{\bf a}_1x^q+\cdots+{\bf a}_{u-1}x^{q^{u-1}}: {\bf a}_j \in {\bf C}_j, j=0,1,\ldots, u-1\},$$ where the matrix at the $i$-th position is the $\bF_q$-linear mapping $a_{0,i}x+\cdots+a_{u-1,i}x^{q^{u-1}}$, ${\bf a}_j=(a_{j,1}, \ldots, a_{j,n})$, $j=0,1,\ldots, u-1$.\\

{\bf Theorem 2.1 (see \cite{Chen1}).} {\em $SR({\bf C}_0, \ldots, {\bf C}_{u-1})$ is a sum-rank code with the block length $n$, the matrix size $u \times u$, the dimension (over ${\bf F}_q$) $u(k_0+\cdots+k_{n-1})$ and the minimum sum-rank distance at least $$d_{sr} \geq \min\{d_0, 2d_1, \ldots, ud_{u-1}\}.$$}

When ${\bf C}_0, \ldots, {\bf C}_{u-1}$ are cyclic codes, then $SR({\bf C}_0, \ldots, {\bf C}_{u-1})$ is a cyclic sum-rank code, see \cite{Cheng}.\\

In the case of sum-rank codes in $\bF_q^{(u,v)} \oplus \cdots \oplus \bF_q^{(u,v)}$, where $u<v$. We identify the matric space ${\bf F}_q^{(u,v)}$ as the space of all $q$-polynomials $a_{u-1}\phi(x^{q^{u-1}})+\cdots+a_0\phi(x)$, where $x \in \bF_{q^u}$, $a_{v-1}, \ldots, a_0 \in {\bf F}_{q^v}$, and $\phi: {\bf F}_{q^u} \longrightarrow {\bf F}_{q^v}$ is a $\bF_q$-linear injective mapping. Let $C_i$ be linear $[n, n-r_i, d_i]_{q^v}$ codes, $i=0,1, \ldots, u-1$. Then we have a linear sum-rank code $$SR({\bf C}_0, \ldots, {\bf C}_{u-1})=\{{\bf a}_{u-1}\phi(x^{q^{u-1}})+{\bf a}_{u-2} \phi(x^{q^{u-2}})+\cdots+{\bf a}_0\phi(x), {\bf a}_j \in {\bf C}_j, j=0,1,\ldots, u-1\}$$ with the dimension (over ${\bf F}_{q^v}$) $$K=un-r_0-\cdots-r_{u-1},$$ and the minimum sun-rank distance $$d_{sr} \geq \min\{d_0, 2d_1, 3d_2, \ldots, ud_{u-1}\}.$$ This is the extension of the construction of sum-rank codes from codes in the Hamming metric given in \cite{Chen1}.\\

\section{Distance-optimal cyclic codes with the minimum distance four}

Let $q$ be a prime power satisfying $q \geq 4$ and $n=\frac{q^m-1}{\lambda}$, where $m=1, 2, 3, \ldots$ are positive integers and $\lambda$ is a divisor of  $q^m-1$. Then each $q$-cyclotomic coset in ${\bf Z}_n$ has at most $m$ elements. The defining set $${\bf T}=C_0 \bigcup C_1 \bigcup C_2$$ has at most $2m+1$ elements. An infinite family of cyclic $[\frac{q^m-1}{\lambda}, \geq \frac{q^m-1}{\lambda}, \geq 4]_q$ codes is constructed, since $0,1,2$ are in the defining set ${\bf T}$.\\

The volume of the ball of the radius $2$ in the Hamming metric space ${\bf F}_q^n$ satisfies that $$V_H(q, 2) > \frac{n(n-1)(q-1)^2}{2}.$$ From the condition $$\lambda <\frac{q-1}{\sqrt{2q(1+\epsilon)}},$$ and $$\lambda \leq q^2,$$ then $$\frac{n(n-1)(q-1)^2}{2}>n(n-1)(1+\epsilon)\geq q^{2m+1},$$ when $m$ is sufficiently large. Then codes in this family are distance-optimal, when $m$ is a sufficiently large positive integer.\\

{\bf Theorem 3.1.} {\em If $\lambda$ is a divisor of $q^m-1$ and satisfies $\lambda <\frac{q-1}{\sqrt{2q(1+\epsilon)}}$, then a distance-optimal cyclic $[\frac{q^m-1}{\lambda}, \frac{q^m-1}{\lambda}-2m-1,4]_q$ code is constructed, when $m$ is sufficiently large.}\\

When $\lambda=1$, we always construct an infinite family of distance-optimal cyclic $[q^m-1, q^m-2-2m, 4]_q$ codes, for any prime power $q$. These distance-optimal cyclic codes have the same parameters as distance-optimal cyclic codes in \cite{YCD,Ding,Wu}.\\

For example, an infinite family of distance=optimal cyclic $[\frac{5^{2m}-1}{3}, \frac{5^{2m}-1}{3}-2m-1, 4]_{25}$ codes can be obtained. It seems this is a family of distance-optimal cyclic codes with new parameters.\\

The Boston bound in \cite{Boston,Zeh} asserts that if $0,1,3,5$ are in the defining set of a $q$-ary cyclic code, then the minimum distance is at least four. Since $3$ is in $C_1$ for a ternary cyclic code and $5$ is in $C_1$ for a quinary cyclic code, we have the following result.\\

{\bf Theorem 3.2.} {\em A distance-optimal ternary cyclic $[3^m-1, 3^m-2m-2,4]_3$ code with the defining set $C_0 \bigcup C_1 \bigcup C_5$ is constructed. A distance-optimal quinary cyclic $[5^m-1, 5^m-2m-2,4]_3$ code with the defining set $C_0 \bigcup C_1 \bigcup C_3$ is constructed.}\\

These two families of distance-optimal codes can be compared with distance-optimal codes in \cite{YCD,Ding,Wu}.\\

\section{Distance-optimal $q$-ary cyclic sum-rank codes with the minimum sum-rank distance four}

We need the following result on the ball of the radius $2$ in the sum-rank metric space.\\

{\bf Lemma 4.1.} {\em There are $\frac{(q^{s_1}-1)(q^{s_2}-1)}{q-1}$ rank one matrices in $\bF_q^{(s_1,s_2)}$.}\\

{\bf Proof.} Each column can be a nonzero vector in ${\bf F}_q^{s_1}$, we have $\frac{q^{s_1}-1}{q-1}$ such nonzero vectors. Moreover, at each column, we can multiply an element in ${\bf F}_q$, this corresponds to a nonzero vector in ${\bf F}_q^{s_2}$. Therefore there are $\frac{(q^{s_1}-1)(q^{s_2}-1)}{q-1}$ rank one $s_1 \times s_2$ matrices. The conclusion follows immediately.\\

{\bf Lemma 4.1.} {\em The volume of the ball $V_{sr}(q, 2) \subset {\bf F}_q^{s \times s} \bigoplus \cdots \bigoplus {\bf F}_q^{s \times s}$ satisfies $$V_{sr}(q, 2) \geq \frac{t(t-1)(q^s-1)^4}{2(q-1)^2}.$$}

{\bf Proof.} This conclusion follows from Lemma 4.1 directly.\\

Let the block length $t$ be $t=\frac{q^{sm}-1}{\lambda}$, where $\lambda$ is a positive divisor of $q^{sm}-1$. The linear cyclic code ${\bf C}_i$ is the trivial $[t, t, 1]_{q^s}$ code, for $i=3, \ldots m-1$. The linear cyclic code ${\bf C}_i$ is the trivial cyclic $[t, t-1, 2]_{q^s}$ code, for $i=1,2$. The first code ${\bf C}_0$ is a linear cyclic $[t, t-2m-1, 4]_{q^s}$ with the defining set ${\bf T}=C_0 \bigcup C_1 \bigcup C_2$. Then the codimension over ${\bf F}_{q^s}$ of this cyclic sum-rank code $SR({\bf C}_0, \ldots, C_{s-1})$ is $2m+3$. \\

{\bf Theorem 4.1.} {\em Let $q$ be a prime power,  $s$ and $m$ be fixed positive integers. Suppose that $\lambda$ is a positive divisor of $q^{sm}-1$ satisfying $$\lambda <\sqrt{\frac{(q^s-1)}{2(q-1)^2(1+\epsilon)}},$$ where $\epsilon$ is an arbitrary small positive real number. Then we construct a cyclic sum-rank code over ${\bf F}_q$ with the block length $t=\frac{q^{sm}-1}{\lambda}$, the matrix size $s \times s$, the cardinality $q^{s^2t-s(2m+3)}$ and the minimum sum-rank distance four. This cyclic sum-rank code is distance-optimal.}\\

{\bf Proof.} From the condition on $\lambda$ and Lemma 4.2, it is easy to verify that $$V_{sr}(q, 2) > q^{s(2m+3)},$$ when $m$ is sufficiently large. The conclusion follows immediately.\\

It is obvious that infinitely many families of distance-optimal cyclic sum-rank codes can be obtained from Theorem 4.1.\\

Let $s_1<s_2$ and $m$ be a positive integers, and $\lambda$ be a divisor of $q^{s_2m}-1$. We construct the block length $t=\frac{q^{s_2m}-1}{\lambda}$, the matrix size $s_1 \times s_2$ distance-optimal sum-rank codes with the minimum sum-rank distance $4$. The linear cyclic code ${\bf C}_i$ is the trivial $[t, t, 1]_{q^{s_2}}$ code, for $i=3, \ldots s_1-1$. The linear cyclic code ${\bf C}_i$ is the trivial cyclic $[t, t-1, 2]_{q^{s_2}}$ code, for $i=1,2$. The first code ${\bf C}_0$ is a linear cyclic $[t, t-2m-1, 4]_{q^{s_2}}$ with the defining set ${\bf T}=C_0 \bigcup C_1 \bigcup C_2$. Then the codimension over ${\bf F}_{q^{s_2}}$ of this cyclic sum-rank code $SR({\bf C}_0, \ldots, C_{s-1})$ is $2m+3$. \\

{\bf Corollary 4.1.} {\em If the divisor $\lambda$ satisfies $$\lambda <\frac{q^{s_1}-1}{q-1}\sqrt{\frac{1}{2(1+\epsilon)q^{s_2}}},$$ then an infinite family of distance-optimal codes of the block length $t$, the matrix size $s_1 \times s_2$ and the minimum distance $4$ is constructed.}\\

We consider the packing density of the distance-optimal cyclic sum-rank code ${\bf C}$  with the block length $t=q^{sm}-1$, the matrix size $s \times s$, the dimension $s^2t-s(2m+3)$ and the minimum distance four, constructed in Theorem 4.1. The volume of the ball with the radius $1$ in the sum-rank metric space ${\bf F}_q^{(s,s)} \oplus \cdots \oplus {\bf F}_q^{(s,s)}$ is $$V_{sr}(q, 1)=1+\frac{(q^s-1)^2}{q-1} t.$$  Then the packing density $$\mu({\bf C})=\frac{V_{sr}(q, 1)}{q^{s(2m+3)}} \approx \frac{1}{q^{sm+s+1}} \approx O(\frac{1}{t}),$$ when $q$ and $s$ are considered as fixed, $m$ goes to the infinity.\\

It is interesting to construct distance-optimal sum-rank codes with minimum sum-rank distance four and larger packing densities.\\

\section{Almost MSRD codes with the block length up to $q^2$}

In this section, we construct $q$-ary almost MSRD codes of the matrix $2 \time 2$ and the minimum sum-rank distance four. The first code ${\bf C}_0$ is the $[t, t-3,4]_{q^2}$ Reed-Solomon code, where $t \leq q^2$, and the second code ${\bf C}_1$ is the trivial $[t,t-1,2]_{q^2}$ code. Then the sum-rank code $SR({\bf C}_0,{\bf C}_1)$ has the dimension (over ${\bf F}_q$) $2(t-1+t-3)=2(2t-4)$ and the minimum sum-rank distance $4$. The Singleton defect is $2(2t-d_{sr}+1)-2(2t-4)=2(2t-3-2t+4)=2$. Then this sum-rank code is almost MSRD code. It is clear for any give block length $t \leq q^2$, these almost MSRD codes with the minimum sum-rank distance four can be constructed explicitly. Comparing MSRD codes with the block length up to $q-1$ constructed in \cite{MP1,Neri}, these almost MSRD codes have larger block lengths up to $q^2$. One of the challenging problem is to construct almost MSRD codes with the block lengths up yo $q^2$ and larger minimum sum-rank distances.\\

\section{Block length $q^4-1$ distance-optimal $q$-ary sum-rank codes with the Singleton defect four}

Let $q$ be a prime power and $n=q^2-1$. We consider the $q$-cyclotomic cosets in ${\bf Z}_n$. Let the defining set be ${\bf T}=C_0 \bigcup C_1 \bigcup C_{q+1}$. Then there are $4$ elements in this defining set. Let ${\bf C}$ be the cyclic code with this defining set. The four elements of this defining set are $0,1, q, q+1$. Set ${\bf A}=\{0,1\}$ and ${\bf B}=\{0, q\}$, then $\gcd(q,n)=1$. The condition of the Hartmann-Tzeng bound in \cite[Theorem 4.5.6]{HP} is satisfied. Then the minimum distance $d({\bf C})$ is at least four. This cyclic $[q^2-1,q^2-5,4]_q$ code is an almost MDS code.\\

Let the block length $t$ be $t=q^4-1$. The linear cyclic code ${\bf C}_1$ is the
trivial $[t, t-1, 2]_{q^2}$ code. The first code ${\bf C}_0$ is a cyclic $[q^4-1, q^4-5, 4]_{q^2}$ code constructed
as above. Then the codimension over ${\bf F}_{q^2}$ of this sum-rank code $SR({\bf C}_0, C_1)$ with the matrix size $2 \times 2$ is $5$. \\

{\bf Theorem 6.1} {\em The $q$-ary and the matrix size $2 \times 2$ sum-rank code $SR({\bf C}_0,{\bf C}_1)$ is distance-optimal. The Singleton defect of this sum-rank code is $4$.}\\

{\bf Proof.} We only need to prove that the volume $V_{sr}(q, 2)$ of the ball with the radius $2$ in the
sum-rank metric space ${\bf F}_q^{(2,2)} \oplus \cdots \oplus {\bf F}_q^{(2,2)}$ satisfies $$V_{sr}(q, 2) >q^{10}.$$
From Lemma 4.1, $V_{sr}(q,2) >\frac{t(t-1)(q^2-1)^4}{2(q-1)^2}>q^{13}$, when $q \geq 3$. The first conclusion is proved.
The Singleton defect is $2(2t-4+1)-(4t-10)=4$. The second conclusion is proved.\\

The above sum-rank codes are cyclic sum-rank codes, see \cite{Cheng}. Infinitely many distance-optimal sum-rank codes, which are close to Singleton-like bound and have much larger $q^4-1>>q-1$ block lengths, are constructed. These sum-rank codes are next best possibility to the almost MSRD codes with the Singleton defect $2$. We do not know if these exists an such almost MSRD code.\\

\section{Quasi-perfect $q$-ary sum-rank codes with the matrix size $2\times m$ and the minimum sum-rank distance three}

We construct quasi-perfect $q$-ary sum-rank codes with the matrix size $2 \times m$ and the block length $t=\frac{q^{mu}-1}{q^m-1}$. The code ${\bf C}_1$ is the trivial $[t, t-1, 2]_{q^m}$ code. The code ${\bf C}_0$ is the Hamming $[t, t-m, 3]_{q^m}$ code. From a similar argument as Theorem 2.1, the minimum sum-rank distance of $SR({\bf C}_0, {\bf C}_1)$ is at least $3$. The codimension of this code over ${\bf F}_{q^m}$ is $m+1$.\\

{\bf Theorem 7.1.} {The sum-rank code $SR({\bf C}_0, {\bf C}_1)$ is distance-optima and quasi-perfect.}\\

{\bf Proof.} The volume of the ball with the radius $2$ is at least $$V_{sr}(q,2) >\frac{t(t-1)}{2} \cdot (q+1)(q^m-1)^2
>q^{m(u+1)}.$$ Then these sum-rank codes are distance-optimal.\\

For any vector ${\bf v}$ in $\bF_q^{(2, m)} \oplus \cdots \oplus \bF_q^{(2,m)}$, it can be represented by ${\bf v}={\bf v}_1 x^q+{\bf v}_0 x$, where ${\bf v}_i \in \bF_{q^m}^t$, $i=0,1$. We considered $\bC_1$ as a covering code with the radius one. It is obvious there is a codeword $\bc_1 \in \bC_1$ such that $\bv_1-\bc_1$ has only one nonzero coordinate. This nonzero coordinate can be at any position in $\{0,1,\ldots, t-1\}$. On the other hand, it is well-known that the Hamming code $\bC_0$ is a perfect code in the Hamming metric. For the vector $\bv_0$ there is a codeword $\bc_0 \in \bC_0$ such that $\bv_0-\bc_0$ has only one nonzero coordinate. This nonzero coordinate has a special coordinate position. Then by choosing $\bc_1$ suitably, we find a codeword $\bc \in SR(\bC_0, \bC_1)$ such that $\bv-\bc$ has only one nonzero coordinate position. Then the sum-rank weight of $\bv-\bc$ is at most $2$. Then the covering radius of the sum-rank code $SR(\bC_0,\bC_1)$ is at most two. It is a quasi-perfect sum-rank code.\\

\section{Quasi-perfect binary sum-rank codes with the matrix size $2 \times 2$ and the minimum distance four}

We recall the following result in \cite{Giulietti}. Let $s_{m,q} = 3(q^{\lfloor\frac{m-3}{2}\rfloor} + q^{\lfloor\frac{m-3}{2}\rfloor-1} + \ldots + q) + 2$ and $n=2q^{\frac{m-2}{2}}+s_{m,q}$. Let $q$ be an even square, $m\geq 7$ be odd. An infinite family of quasi-perfect $[n,n-m, 4]_{q}$ codes was constructed, see \cite[Proposition 2.5]{Giulietti}. Other infinite families of quasi-perfect $[n, n-m, 4]_{q}$ codes was constructed in \cite[Propositon 4.1]{Giulietti}, for even prime power $q$. It is obvious that quasi-perfect codes and sum-rank codes with even minimum distances and even minimum sum-rank distances are distance-optimal.\\

We need the following lemma for binary sum-rank codes of the matrix size $2 \times 2$.\\

{\bf Lemma 8.1.} {\em Let ${\bf C}_0 \subset {\bf F}_4^t$ and ${\bf C}_1 \subset {\bf F}_4^t$ be two linear $[t, k_0, d_0]_4$ and $[t, k_1, d_1]_4$ codes over ${\bf F}_4$. Then a block length $t$ and matrix size $2 \times 2$ binary linear sum-rank-metric code $SR({\bf C}_0, {\bf C}_1)$ can be constructed explicitly. The minimum sum-rank distance of $SR({\bf C}_0, {\bf C}_1)$ is at least $\min\{d_0, 2d_1\}$. The dimension of $SR({\bf C}_0, {\bf C}_1)$ over ${\bf F}_2$ is $\dim_{{\bf F}_2}(SR({\bf C}_0, {\bf C}_1))=2(k_0+k_1)$. Moreover for any two codewords, ${\bf a}_0 \in {\bf C}_0$ and ${\bf a}_1 \in {\bf C}_1$, set $$I=supp({\bf a}_0) \cap supp({\bf a}_1),$$ then $$wt_{sr}({\bf a}_1x+{\bf a}_0x^2)=2wt_H({\bf a}_0)+2wt_H({\bf a}_1)-3|I|.$$}\\

{\bf Proof.} For the coordinate position $i \in supp({\bf a}_0)\backslash supp({\bf a}_1)$, it is clear the matrix ${\bf a}_1x+{\bf a}_0x^2$ in this position is of the rank $2$. For the coordinate position $i \in supp({\bf a}_1)\backslash supp({\bf a}_0)$, it is clear the matrix ${\bf a}_1x+{\bf a}_0x^2$ in this position is of the rank $2$. For these coordinate positions $i \in I$, it is clear that the matrix is of rank $1$, since there is one nonzero root $x=\frac{a_{i,1}}{a_{i,0}}$, where $${\bf a}_0=(a_{1,0}, \ldots, a_{t,0}),$$ $${\bf a}_1=(a_{1,1}, \ldots, a_{t,1}).$$ Then it follows that $$wt_{sr}({\bf a}_1x+{\bf a}_0x^2)=2(wt_H({\bf a}_0)-|I|)+2(wt_H({\bf a}_1)-|I|)+|I|.$$ The conclusion follows immediately.\\

From Lemma 8.1, for two vectors ${\bf a}_0$ and ${\bf a}_1$ in $\bF_4^t$ satisfying $$supp({\bf a}_0)=supp({\bf  a}_1),$$ $$wt_{sr}({\bf a}_1x+{\bf a}_0x^2)=wt({\bf a}_0)=wt({\bf a}_1).$$

From the following result, quasi-perfect binary sum-rank codes of the matrix size $2 \times 2$ can be obtained.\\

{\bf Theorem 8.1.} {\em Let ${\bf C}_1 \subset {\bf F}_4^t$ be the linear $[t, t-1,2 ]_4$ and $[t, k_2, 4]_4$ code and  and ${\bf C}_0 \subset {\bf F}_4^t$ be  a $[t, k, 4]_4$ code with the covering radius two.  Then a block length $t$ and matrix size $2 \times 2$ binary linear sum-rank-metric code $SR({\bf C}_0, {\bf C}_1)$ with the minimum sum-rank distance four is constructed explicitly. The covering radius of this binary sum-rank code is two. Then $SR({\bf C}_0, {\bf C}_1)$ is quasi-perfect.}.\\

{\bf Proof.} For any vector ${\bf v}$ in $\bF_4^{(2, 2)} \oplus \cdots \oplus \bF_q^{(2,2)}$ ($t$ copies), it can be represented by ${\bf v}={\bf v}_1 x^2+{\bf v}_0 x$, where ${\bf v}_i \in \bF_{4}^t$, $i=0,1$. We considered $\bC_1$ as a covering code with the radius one. It is obvious there is a codeword $\bc_1 \in \bC_1$ such that $\bv_1-\bc_1$ has exactly two nonzero coordinates. These two nonzero coordinates can be at any two different positions in $\{0,1,\ldots, t-1\}$. On the other hand, the quasi-perfect code $\bC_0$ has the covering radius two in the Hamming metric. For the vector $\bv_2$ there is a codeword $\bc_0 \in \bC_0$ such that $\bv_0-\bc_0$ has at most two nonzero coordinates. These two nonzero coordinates have special coordinate positions. Then by choosing $\bc_1$ suitably, we find a codeword $\bc \in SR(\bC_0, \bC_1)$ such that $\bv-\bc$ has at most two nonzero coordinate positions. Then the sum-rank weight of $\bv-\bc$ is at most $2$. Then the covering radius of the binary sum-rank code $SR(\bC_0,\bC_1)$ is at most two. It is a quasi-perfect sum-rank code.\\

From Theorem 8.1, many quasi-perfect binary sum-rank codes with the matrix size $2 \times 2$ and the minimum sum-rank distance four can be constructed explicitly, from these quasi-perfect codes over $\bF_4$ constructed in \cite{Giulietti}.\\

\section{Plotkin sum of sum-rank codes}

As in the case of the matrix size $1 \times 1$ sum-rank codes (codes in the Hamming metric), the Plotkin sum of sum-rank codes of the matrix size $n \times m$, $n\leq m$,  can be defined and nice sum-rank codes can be obtained.\\

Let ${\bf C}_1$ and ${\bf C}_2$ be two sum-rank codes in $\bF_q^{(n,m)} \oplus \cdots \oplus \bF_q^{(n,m)}$, with minimum sum-rank distances $d_1$, $d_2$, and dimensions $k_1$ and $k_2$. Then we define their Plotkin sum-rank code as $$\Plotkin({\bf C}_1, {\bf C}_2)=\{({\bf c}_1|{\bf c}_1+{\bf c}_2): \bc_1 \in  \bC_1, \bc_2 \in \bC_2\}.$$\\

{\bf Theorem 9.1.} {\em The dimension of the Plotkin sum is $k_1+k_2$ and the minimum sum-rank distance the Plotkin sum is $\min\{2d_1, d_2\}$.}\\

{\bf Proof.} If ${\bf c}_2$ is not zero, $$wt_{sr}({\bf c}_1|{\bf c}_1+{\bf c}_2)=wt_{sr}(\bc_1)+wt_{sr}(\bc_1+\bc_2)\geq wt_{sr}(\bc_1+(\bc_1+\bc_2))=wt_{sr}(\bc_2)\geq d_2.$$ If $\bc_2=0$, then $$wt_{sr}({\bf c}_1|{\bf c}_1)=2wt_{sr}(\bc_1) \geq 2d_1.$$
The conclusion follows immediately.\\

From Theorem 2.1, we get the block length $t$ sum-rank code ${\bf C}_1$ with the matrix size $s \times $ and the minimum sum-rank distance $2$ and the codimension $s$ directly, for an arbitrary positive integer $t$. From Theorem 4.1, we get an infinite family of distance-optimal binary sum-rank codes $\bC_2$ with the block length $t=2^{sm}-1$, the matrix size $s \times s$, the codimension $2m+3$ (over $\bF_{2^s}$), and the minimum sum-rank distance four.\\

Then we can obtain new distance-optimal binary sum-rank codes with the block length $2t=2(2^{sm}-1)$.\\

{\bf Corollary 9.1.} {\em Let $s$ and $m$ be two positive integers. Then the Plotkin sum $\Plotkin(\bC_1, \bC_2)$ is another distance-optimal binary sum-rank codes with the block length $2t$.}\\

{\bf Proof.} The minimum distance of $\Plotkin(\bC_1, \bC_2)$ is four. $$V_{sr}(2, 2) >\frac{2t(2t-1)}{2}\cdot (2^s-1)^2= 2(2^{sm}-1)(2^{sm}-\frac{3}{2})(2^s-1)^2 \geq 2^{s(2m+4)},$$ when $m$ is large. The conclusion follows directly.\\

\section{Conclusion}

To construct distance-optimal and quasi-perfect codes in the sum-rank metric are interesting and challenging problems. In this paper, infinite many families of distance-optimal cyclic sum-rank codes with the matrix size $2\times 2$ and the minimum sum-rank distance four were constructed.  An infinite family of long distance-optimal $q$-ary sum-rank codes with the block length $q^4-1$, the matrix size $2 \times 2$, the minimum sum-rank distance four and the Singleton defect four was given. Moreover, several families of quasi-perfect sum-rank codes with the matrix size $n \times m$, $2 \leq n \leq m$, were also presented. We also constructed almost MSRD codes with the block length up to $q^2$ and showed that the Plotkin sum of sum-rank codes gives us more distance-optimal binary sum-rank codes.\\


\begin{thebibliography}{10}



\bibitem{AKR} A. Abiad, A. Khramova and A. Ravagnani, Eigenvalue bounds for sum-rank-metric codes,
IEEE Trans. Inf. Theory, early access, 2023.

\bibitem{Boer} M. A. de Boer, Almost MDS codes, Des., Codes and Cryptogr., vol. 9. pp. 143-155, 1996.



\bibitem{BBDF} T. Baicheva, I. Bouyukliev, S. Dodunekov and V. Fack, Binary and ternary linear quasi-perfect codes with small dimensions, IEEE Trans. Inf. Theory, vol. 54, no. 9, pp. 4335-4339, 2008.


\bibitem{BC1} R. C. Bose and D. K. Ray-Chaudhuri, On a class of error-correcting binary group codes, Inf. and Contr., vol. 3, pp. 68-79, 1960.

\bibitem{BC2} R. C. Bose and D. K. Ray-Chaudhuri, Further results on error-correcting binary group codes, Inf. and Contr., vol. 3, pp. 279-290, 1960.



\bibitem{Boston} N. Boston, Bounding minimum distances of cyclic codes using algebraic geometry, Electron. Notes Discr. Math., vol. 6, pp. 385-394, 2001.

\bibitem{BGR} E. Byrne, H. Gluesing-Luerssen and A. Ravagnani, Fundamental properties of sum-rank-metric codes, IEEE Trans. Inf. Theory, vol. 67, no. 10, pp. 6456-6475, 2021.


\bibitem{Chen1} H. Chen, New explicit linear sum-rank-metric codes, IEEE Trans. Inf. Theory, vol. 69, no. 10, pp. 6303-6313, 2023.

\bibitem{Chen} H. Chen, Many non-Reed-Solomon type MDS codes from arbitrary genus algebraic curves, IEEE Trans. Inf. Theory, early access, 2023.


\bibitem{Cheng} Z. Cheng, C. Xie, H. Chen and C. Ding, Cyclic and negacyclic sum-rank codes, arXiv:2401.04885, 2024.




\bibitem{CHLL} G.  D. Cohen, I. Honkala, S. Litsyn and A. Lobstein, Covering codes, North-Hollan Math, Libarary, Elsecier, 1997.

\bibitem{Ding} C. Ding and T. Helleseth, Optimal ternary cyclic codes from monomials, IEEE Trans. Inf. Theory, vol. 59, no. 9, pp. 5898-5904, 2013.

\bibitem{DingLi} C. Ding and C. Li, BCH cyclic codes, submitted, 2023.

\bibitem{Dodun} S. Dodunekov, Some quasi-perfect double error correcting codes, Probl. Control. Inform. Theory, vol. 15, no. 5, pp. 367-375, 1986.



\bibitem{Etzion} T. Etzion and B. Mounits, Quasi-perfect codes with small distance, IEEE Trans. Inf. Theory, vol. 51,  no. 11, pp. 3938-3946, 2005.

\bibitem{Etzion1} T. Etzion, Perfect codes and related structures, World Scientific, Singapore, 2022.


\bibitem{Giulietti} M. Giulietti and F. Pasticci, Quasi-perfect linear codes with minimum distance 4, IEEE Trans. on Inf. Theory, vol. 53,  no. 5, pp. 1928-1934, 2007.


\bibitem{Heng} Z. Heng, C. Ding and W. Wang, Optimal binary linear codes from maximal arcs, IEEE Trans. Inf. Theory, vol. 66, no. 9, pp. 5387-5394, 2020.


\bibitem{HF} D. Han and C. Fan, Roth-Lemple NMDS codes of non-elliptic-curve type, IEEE Trans. Inf. Theory, vol. 69, no. 9, pp. 5670--5675, 2023.

\bibitem{Heng1} Z. Heng, Q. Wang and C. Ding, Two families of optimal linear code and their subfield codes, IEEE Trans. Inf. Theory, vol. 66, no. 11, pp. 6872-6883, 2020.


\bibitem{Hoc} A. Hocquenghem, Codes correcteurs d'erreurs, Chiffres (Paris), vol. 2, pp. 147-156, 1959.


\bibitem{Hu} Z. Hu, N. Li, X. Zeng, L. Wang and X. Tang, The subfield-based construction of optimal linear codes over finite fields, IEEE Trans. Inf. Theory, vol. 68, no. 7, pp. 4408-4421, 2023.

\bibitem{HP} W. C. Huffman and V. Pless, Fundamentals of error-correcting codes, Cambridge University Press, Cambridge, U. K., 2003.







\bibitem{LiHell} C. Li and T. Helleseth, Quasi-perfect linear codes from planar and APN functions, Cryptogra. Commun., vol. vol. 8, pp. 215-227, 2016.

\bibitem{Lu} H. -F. Lu and P. V. Kumar, A unified construction of space-time codes with optimal rate-diversity tradeoff. IEEE Trans. Info. Theory, vol. 51, no. 5, pp. 1709-1730, 2005.


\bibitem{MScode} F. J.  MacWilliams and N. J. A. Sloane, The Theory of error-correcting codes, 3rd Edition, North-Holland Mathematical Library, vol. 16. North-Holland, Amsterdam, 1977.

\bibitem{MP1} U. Mart\'{\i}nez-Pe\~{n}as, Skew and linearized Reed-Solomon codes and maximal sum rank distance codes
over any division ring, Jour. Alge., vol. 504, pp. 587-612, 2018.

\bibitem{MK} U. Mart\'{\i}nez-Pe\~{n}as and F. R. Kschischang, Universal and dynamic locally repairable codes with maximally recoverablity via sum-rank codes, IEEE Trans. Inf. Theory, vol. 65,  no. 12, pp. 7790-7805, 2019.
		

\bibitem{MP191} U. Mart\'{\i}nez-Pe\~{n}as, Hamming and simplex codes from the sum-rank metic, Des., Codes and Cryptogr., vol. 88, pp. 1521-1539, 2019.

\bibitem{MK19} U. Mart\'{\i}nez-Pe\~{n}as and F. R. Kschischang, Reliable and secure multishot network coding using linearized Reed-Solomon codes, IEEE Trans. Inf. Theory, vol. 65,  no. 8, pp. 4785-4803, 2019.

\bibitem{MP21} U. Mart\'{\i}nez-Pe\~{n}as, Sum-rank BCH codes and cyclic-skew-cyclic codes, IEEE Trans. Inf. Theory, vol. 67,  no. 8, pp. 5149-5167, 2021.	

\bibitem{MPK22} U. Mart\'{\i}nez-Pe\~{n}as, M. Shehadeh and F. R. Kschischang, Codes in the sum-rank metric, Foundamentals and applications, Foundations and Trends in Communications and Information Theory, vol. 19, no. 5, pp. 814-1031, 2022.


\bibitem{MPP} U. Mart\'{\i}nez-Pe\~{n}as and S. Puchinger, Maximum sum-rank distance codes over finite chain rings, IEEE Trans. Inf. Theory, early access, 2024.

\bibitem{MP24} U. Mart\'{\i}nez-Pe\~{n}as, New constructions of MSRD codes, arXiv:2402.03084,2024.

\bibitem{Neri} A. Neri, Twisted linearized Reed-Solomon codes: A skew polynomial framework, 2021,
Jour. Alge., vol. 609, pp. 792-839, June, 2022.

\bibitem{NSZ21} A. Neri, P. Santonastaso and F. Zullo, The geometry of one-weight codes in the sum-rank metric, Jour. Combin. Theory, Ser.A, vol. 194, 105703, 2023.


\bibitem{NPS} D. Napp, R. Pinto and V. Sidorenko, Concatenation of covolutional codes and rank metric codes for muti-shot network coding, Des., Codes and Cryptogra., vol. 86, no. 2, pp. 303-318, 2018.


\bibitem{NU} R. W. Nobrega and B. F. Uchoa-Filho, Multishot codes for network coding using rank-metric codes, 3rd IEEE Intentional Workshop on Wireless Network Coding, June, 2010.


\bibitem{Prange} E. Prange, Cyclic error-correcting codes in two symbols, TN-57-013, Technical notes issued by Air Force Cambridge Research Labs, 1957.




\bibitem{SK} M. Shehadeh and F. R. Kschischang, Space-time codes from sum-rank codes, IEEE Trans. Inf. Theory, vol. 68, no. 3, pp. 1614-1637, 2022.


\bibitem{Lint} J. H. van Lint, Introduction to the coding theory, GTM 86, Third and Expanded Edition, Springer, Berlin, 1999.


\bibitem{Wagner} T. Wagner, A search technique for quasi-perfect codes, Inf. and Contr., vol. 9, pp. 94-99, 1966.


\bibitem{Wang} X. Wang, D. Zheng and C. Ding, Some punctured codes of sveral families of binary linear codes, IEEE Trans. Inf. Theory, vol. 67, no. 8, pp. 5133-5148, 2021.


\bibitem{Wu} G. Wu, H. Liu and Y. Zhang, Several classes of optimal $p$-ary cyclic codes with minimum distance four,
 Finite Fields Appl., vol. 92, 102275, 2023.


\bibitem{Xiong} M. Xiong and N. Li, Optimal cyclic codes with generalized Niho-type zeros and weight distributions, IEEE Trans. Inf. Theory, vol. 61, no. 9, pp. 4914-4922, 2015.


\bibitem{YCD} J. Yuan, C. Carlet and C. Ding, The weight distribution of a class
of linear codes from perfect nonlinear functions, IEEE Trans. Inf.
Theory, vol. 52, no. 2, pp. 712-717, 2006.

\bibitem{Zeh} A. Zeh, A. Wachter-Zeh and S. Bezzateev, Decoding cyclic codes up to a new bound on the minimum distance, IEEE Trans. Inf. Theory, vol. 58, no. 6, pp. 3951-3960, 2012.





\end{thebibliography}
\end{document}